\documentclass[prb,aps,amsmath,twocolumn,amssymb,floatfixng,
superscriptaddress]{revtex4}

\usepackage[dvips]{graphics}
\usepackage{color}
\definecolor{dred}{rgb}{0,0.75,0.5}

\begin{document}

\title{\textcolor{dred}{Selective spin transport through a quantum 
heterostructure: Transfer matrix method}}

\author{Moumita Dey}

\affiliation{Department of Physics, Adamas University, Barasat-Barrackpore 
Road, Jagannathpur, Kolkata-700 126, India}

\author{Santanu K. Maiti}

\email{santanu.maiti@isical.ac.in}

\affiliation{Physics and Applied Mathematical Unit, Indian Statistical
Institute, 203 Barrackpore Trunk Road, Kolkata-700 108, India}

\begin{abstract}
In the present work we propose that a one-dimensional quantum heterostructure 
composed of magnetic and non-magnetic atomic sites can be utilized as a spin 
filter for a wide range of applied bias voltage. A simple tight-binding 
framework is given to describe the conducting junction where the 
heterostructure is coupled to two semi-infinite one-dimensional non-magnetic 
electrodes. Based on transfer matrix method all the calculations are 
performed numerically which describe two-terminal spin dependent
transmission probability along with junction current through the wire. 
Our detailed analysis may provide fundamental aspects of selective spin 
transport phenomena in one-dimensional heterostructures at nano-scale level.
\end{abstract}

\maketitle

\section{Introduction}

Rapid progress of nanoscience and nanotechnology has allowed us to 
establish astonishing magneto devices which reveal several enchanting 
phenomena~\cite{r1}. This new span in the study of magnetism has been 
instigated in 1988 after the discovery of Giant Magneto Resistance (GMR) 
effect~\cite{r2} detected in magnetic multilayers formed by alternating 
magnetic and non-magnetic (NM) materials. 

\begin{figure}[ht]
{\centering \resizebox*{7cm}{5cm}{\includegraphics{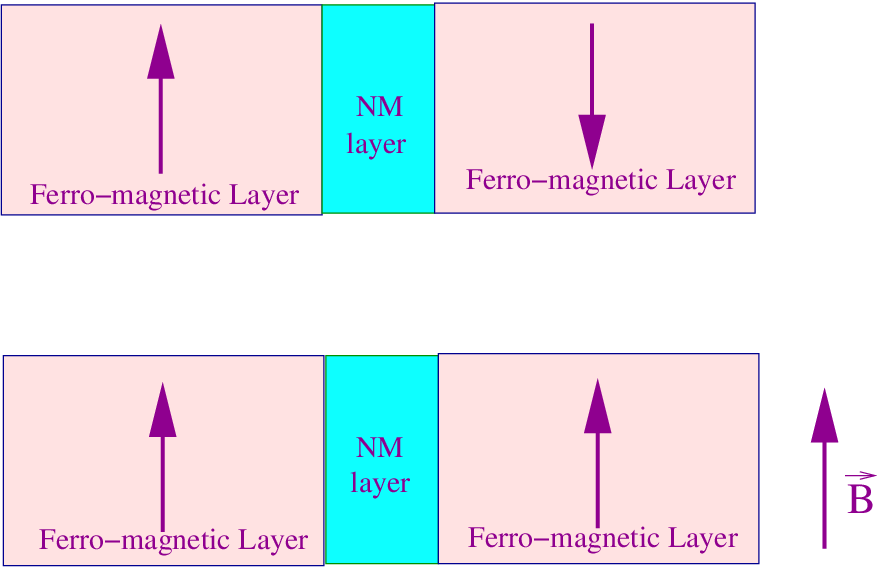}}\par}
\caption{(Color online). Schematic representation of a GMR sensor.}
\label{figure4}
\end{figure}

It is well known that in the absence of an external magnetic field, 
exchange coupling between neighboring magnetic layers through the 
non-magnetic one aligns the magnetization vector anti-parallel to each 
other. Then, when a strong magnetic field which is enough to overcome 
the anti-ferromagnetic coupling is applied, all the magnetization vectors 
orient along the field direction. This new parallel configuration yields 
an electrical resistance which is much smaller compared to the 
anti-ferromagnetic configuration. This substantial change in resistance 
is known as the GMR effect which is demonstrated in Fig.~\ref{figure4}. 
We can define GMR mathematically as, 
\begin{equation}
R=\frac{R_{AP} - R_P}{R_P}
\label{equation49}
\end{equation}
where the symbols have their usual meanings. In conventional GMR effect, 
$R_{AP} > R_P$, and $R$ is unbounded. Another definition can also be given 
for GMR which is,
\begin{equation}
R^{\prime}=\frac{R_{AP} -R_P }{R_{AP}},
\label{equation50}
\end{equation}
where, $0\leq{R^{\prime}}\leq1$.

Along with this, the inverse GMR effect has also been discovered in some 
materials for which $R_P > R_{AP}$. The study of GMR effect is of underlying 
importance as it confirms the fact that the spin of an electron can also 
take a salient role in transport phenomena. In contrast with the past, the 
study of GMR effect has drawn much curiosity from academic circuit to 
commercial levels~\cite{r3} as GMR based magnetic data storage devices are 
used in almost all computers. Not only that, spintronics in low-dimensional 
systems furnishes some tangible advantages over bulk metals and 
semiconductors. In molecular systems, the traditional mechanisms for 
spin decoherence (spin-orbit coupling, scattering from paramagnetic 
impurities, etc.) get reduced. Hence, in nano-scale systems, we can 
anticipate the spin coherence time to be several orders of magnitude 
larger than in bulk systems. Therefore, the study of spin dependent 
transport and spin dynamics is of pronounced importance to understand 
and to flourish the field, {\em spintronics}.

Spin dependent transport through mesoscopic systems has attracted affluent
attention in the evolution of spintronics~\cite{r4,r5,r6,r7,skm1,skm2,
skm3,skm4,skm44}. Several experimental attempts have been done to study spin 
transport in quantum dots (QD)~\cite{r8,r9,r10,r11} and other molecular 
systems~\cite{r11,r12,r13,r14}.
In these systems, electrical resistance depends on the spin state 
of electrons passing through the device, and it can be steered by an 
applied magnetic field. This is because of the misproportion in 
transmission probability for spin up and down electrons through the 
device~\cite{r15,r16}, and, it can be utilized to achieve efficient spin
filter, the design of which is of great concern in the present era of 
nanofabrication. Though much effort has been devoted to realize efficient
spin filter and high degree of spin polarization, but none of them is quite
suitable from experimental perspective. Therefore, it is major challenge 
to propose a {\em simple} model that can generate pure spin current and 
provide efficient spin filter operation, and also can be constructed 
through simple experimental set-up. Device with magnetic quantum dots 
(MQDs) can provide a possible route towards this direction.

The spin dependent transport through such a QD system can be inspected 
by coupling it to external electrodes and passing a current through the 
system~\cite{r14,r17,r18,r19,r20}. Several theoretical as well as 
experimental studies have been made so far on spin transport through 
quantum dot devices~\cite{r21,r22,r23}. Aim of our present model is 
to study coherent spin transport through a quantum heterostructure 
attached to two non-magnetic electrodes. Within a simple tight-binding 
framework we numerically calculate two-terminal spin dependent transmission 
probability together with junction current based on transfer matrix 
method~\cite{r24,r25,r26}. Several cases are analyzed depending on the
orientations of magnetic and non-magnetic sites of quantum heterostructure 
and our results suggest that under certain conditions the model can be 
utilized as an efficient spin filter with high degree of spin polarization 
($100\%$) for a wide range of applied bias voltage.

The paper is organized as follows. In Sec. II, the model and the methods 
for the calculations of spin dependent transmission probabilities along 
with transport currents are described. In Sec. III we present the numerical
results of heterostructure considering four different configurations and
analyze how to achieve spin selective transmission and polarized spin 
currents. Finally, in Sec. IV we conclude our findings.

\section{Description of the model and theoretical formalism}

\subsection{The Model and the Hamiltonian}

The model chosen is an array of quantum dots formed by a sequence of magnetic 
and non-magnetic sites connected to two non-magnetic leads, viz, source and
\begin{figure}[ht]
{\centering \resizebox*{7.5cm}{1.7cm}{\includegraphics{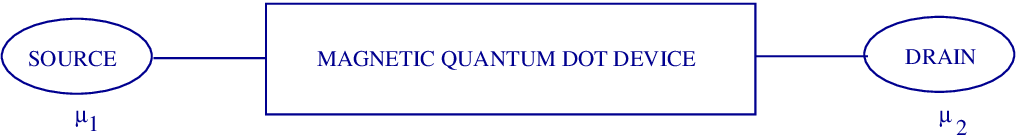}}\par}
\caption{(Color online). Schematic view of magnetic quantum dot device 
(magnetic spacer) coupled to two leads, source and drain.}
\label{figure5}
\end{figure}
drain. The spin dependent electronic transport through this structure is 
studied including the spin flip scattering effect. The bridging conductor 
is basically a repetition of a chosen unit cell formed by magnetic and/or
non-magnetic atoms. The geometry of the bridge structure is schematically 
shown in Fig.~\ref{figure5} which is further clearly viewed from 
\ref{figure6}.
The direction of magnetization on each magnetic site in the MQD device is 
chosen to be arbitrary, specified in each site $n$, which in spherical 
\begin{figure}[ht]
{\centering \resizebox*{7.5cm}{2cm}{\includegraphics{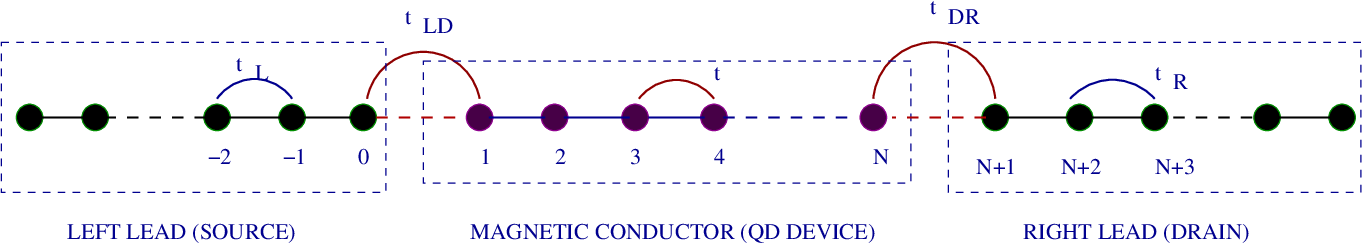}}\par}
\caption{(Color online). Schematic view of the one-dimensional bridge system 
where magnetic conductor is connected to non-magnetic source and drain.}
\label{figure6}
\end{figure}
polar coordinate system is defined by two angles $\theta_n$ and $\phi_n$. 
Here $\theta_n$ is the angle between magnetization direction and the $Z$-axis,
$\phi_n$ is the azimuthal angle of magnetization measured from $X$-axis at 
\begin{figure}[ht]
{\centering \resizebox*{7cm}{7cm}{\includegraphics{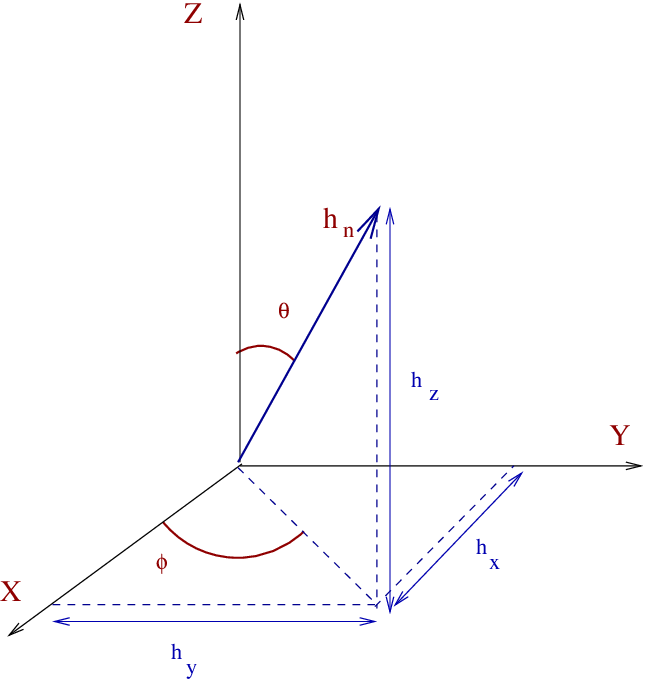}}\par}
\caption{(Color online). Decomposition of $\vec{\bf h}_n$ in spherical 
polar coordinates.}
\label{figure7}
\end{figure}
site $n$ as shown in Fig.~\ref{figure7}.
The whole system can be described by a Hamiltonian like, 
\begin{equation}
H=H_L + H_{LD} + H_D + H_{DR} + H_R
\label{equation51}
\end{equation}
The spin polarized electrons for a $N$-site quantum dot can be described 
within the effective one-electron picture in tight-binding framework and 
the nearest-neighbor approximation as,
\begin{eqnarray}
H_D & = & \sum_{n}{\bf c_{n}^{\dagger}} \left({\bf \epsilon}_{n} - 
\vec{\bf{h}_n}.\vec{\bf{\sigma}}\right)\bf c_{n} + 
\sum_{n}\left(\bf c^{\dag}_{n}
\bf t_{n,n+1}\bf c_{n+1} \right. \nonumber \\
 & & \left. + \bf c^{\dag}_{n+1}\bf t_{n,n+1}\bf c_{n}\right) 
\label{equation52}
\end{eqnarray}
where,
$\bf c_{n}^\dag =
\left( \begin{array}{cc}
c_{n\uparrow}^{\dag} & c_{n\downarrow}^{\dag}
\end{array} \right)$,
~\\
$\bf c_{n} = 
\left( \begin{array}{c}
c_{n\uparrow} \\
c_{n\downarrow}
\end{array} \right)$,
~\\
${\bf \epsilon_{n}} = 
\left( \begin{array}{cc}
\epsilon_{n\uparrow} & 0 \\
0 & \epsilon_{n\downarrow}
\end{array} \right)$,
~\\
$\bf t_{n,n+1} = 
\left( \begin{array}{cc}
t_{n,n+1} & 0 \\
0 & t_{n,n+1}
\end{array} \right)$
\vskip 0.25cm
\noindent
The term $\bf{\epsilon_n}$ of Eq.~\ref{equation52} describes the on-site 
energy. The term $\vec{\bf{h}_n}.\vec{\bf{\sigma}}$ describes the 
interaction of 
electrons with the magnetic atoms in MQD device. This term allows the 
spin flips at the magnetic sites. $h_n$ is the amplitude of the spin 
flip parameter at site $n$. $\bf{\sigma}$ is the Pauli spin operator having 
components $\bf{(\sigma_x,\sigma_y,\sigma_z)}$. Spin flip scattering is 
dependent on the magnetic moment orientation of the atoms in MQD device 
with respect to the $Z$-axis. The term $t_{n,n+1}$ is the nearest-neighbor 
hopping integral, $c_{n\sigma}^{\dagger}(c_{n\sigma})$ is the 
creation(annihilation) operator for an electron on site $n$ with
spin $\sigma$. The Hamiltonian of the left(right) lead is defined as, 
\begin{equation}
H_{L(R)} = \sum_i {\bf c_i^{\dagger} \epsilon_i c_i} + \sum_i 
\left(\bf c^{\dag}_i \bf t_{L(R)} \bf c_{i+1} +
\bf c^{\dag}_{i+1} \bf t_{L(R)} \bf c_{i}\right)
\label{equation53}
\end{equation}
where $t_{L(R)}$ refers to the hopping integral between the sites of 
the left(right) lead. Finally, the Hamiltonian that corresponds to the 
coupling of the MQD device to the leads can be expressed in the form,
\begin{eqnarray}
H_{LD(RD)} & = & \left(\bf c^{\dag}_{0(N+1)} \bf t_{LD(RD)} \bf c_{1(N)}
\right.\nonumber \\
 && \left. +\bf c^{\dag}_{1(N)} \bf t_{LD(RD)} \bf c_{0(N+1)}\right)
\label{equld}
\end{eqnarray}
Throughout this study, it has been assumed that the nonmagnetic leads 
are ideal, i.e., their resistances have been neglected. The main 
contribution to the resistance in this ballistic device has come from 
contact resistance and spin scattering in MQD. 

Using this model spin dependent transmission coefficients and junction
currents are calculated to investigate the transport properties following 
transfer matrix approach. In this framework Schr\"{o}dinger equation is 
written in terms of $\psi_{n,\sigma}$ (localized Wannier basis), which 
express the amplitude of wave function at site $n$, with energy $E$ and
spin $\sigma$.

\subsection{Transfer matrix method}

In order to calculate spin dependent transmission probabilities, the
Schr\"{o}dinger equation for the electron wave function must be solved 
in the MQD device. So the starting point is the time independent 
Schr\"{o}dinger equation in the MQD device, which can be written as,
\begin{equation}
H|\phi\rangle = E|\phi\rangle
\label{equation55}
\end{equation} 
where,
\begin{equation}
|\phi \rangle = \sum_i [ \psi_{i\uparrow} |i\uparrow\rangle +
\psi_{i\downarrow} |i\downarrow\rangle ]
\label{equation56}
\end{equation} 
Here, $|\phi\rangle$ is written as a linear combination of spin up and 
spin down Wannier states. Now, 
\begin{eqnarray}
\vec h.\vec \sigma & = & h_x\sigma_x + h_y\sigma_y + h_z\sigma_z \nonumber \\
& = &
\left( \begin{array}{cc}
h \cos\theta_n & h \sin\theta_n e^{-i\phi_n} \\
h \sin\theta_n e^{i\phi_n} & -h \cos \theta_n 
\end{array} \right)
\label{equation57}
\end{eqnarray}
So the Hamiltonian $H_D$ for the whole MQD device can be written as,
\begin{widetext}
\begin{eqnarray}
H_D & = & \sum_n \left(c^{\dag}_{n\uparrow} c^{\dag}_{n\downarrow}\right)
\left(\begin{array}{cc}
\epsilon_{n\uparrow}-h \cos\theta_n & -h \sin\theta_n e^{-i\phi_n} \\
-h \sin\theta_n e^{i\phi_n} & \epsilon_{n\downarrow}+h \cos \theta_n 
\end{array} \right)
\left( \begin{array}{c}
c_{n\uparrow} \\
c_{n\downarrow} \end{array} \right) + \nonumber \\
& &\sum_n\left[\left(c{^\dag}_{n\uparrow} c{\dag}_{n\downarrow}\right) 
\left( \begin{array}{cc}
t & 0 \\
0 & t \end{array} \right)
\left( \begin{array}{c}
c_{n+1,\uparrow} \\
c_{n+1,\downarrow} \end{array} \right)  
+ \left(c{^\dag}_{n+1,\uparrow} c{\dag}_{n+1,\downarrow}\right) 
\left( \begin{array}{cc}
t & 0 \\
0 & t \end{array} \right)
\left( \begin{array}{c}
c_{n,\uparrow} \\
c_{n,\downarrow} \end{array} \right)\right] \nonumber\\
& = & H_1 + H_2   
\label{equation58}
\end{eqnarray}
\end{widetext}
where,
\begin{eqnarray}
H_1 & = & \sum_n|n\uparrow \rangle \left(\epsilon_{n\uparrow} -
h \cos\theta_n \right) 
\langle n\uparrow| \nonumber \\
& -& \sum_n|n\uparrow \rangle h \sin\theta_n e^{-i\phi_n} 
\langle n\downarrow| \nonumber \\
& - & \sum_n|n\downarrow \rangle h \sin\theta_n e^{i\phi_n} \langle n\uparrow| 
\nonumber \\
& + & \sum_n|n\downarrow \rangle \left(\epsilon_{n\downarrow} + 
h \cos\theta_n\right) \langle 
n\downarrow|
\label{equation59}
\end{eqnarray}
\begin{eqnarray}
H_2 & = & \sum_n|n\uparrow \rangle t \langle n+1,\uparrow| + 
\sum_n|n\downarrow \rangle t \langle n+1,\downarrow| \nonumber \\ 
& + & \sum_n|n+1,\uparrow \rangle t \langle n \uparrow| + 
\sum_n|n+1,\downarrow \rangle t \langle n \downarrow| \nonumber \\
\label{equation60}
\end{eqnarray}
Operating $H$ on $|\phi \rangle$ we get the following two equations 
relating the Wannier amplitudes on site $n$ of the MQD device with the
neighboring $n\pm 1$ sites,
\begin{eqnarray}
\left(E-\epsilon_n + h_n \cos \theta_n \right) \psi_{n\uparrow} + 
h_n \sin\theta_n e^{-i\phi_n} \psi_{n\downarrow} \nonumber \\
= t \psi_{n+1,\uparrow} +
t \psi_{n-1,\uparrow}
\label{equation61}
\end{eqnarray}
\begin{eqnarray}
h_n \sin\theta_n e^{i\phi_n} \psi_{n\uparrow} + \left(E-\epsilon_n-h_n 
\cos\theta_n \right) \psi_{n\downarrow} \nonumber \\
= t \psi_{n+1,\downarrow} + 
t \psi_{n-1,\downarrow}
\label{equation62}
\end{eqnarray}
Transfer matrix for the $n$th site relates the wave amplitudes of $n$th 
site with that of $(n-1)$th and $(n+1)$th sites. So we can write the 
transfer matrix equation for the $n$th site as,
\begin{eqnarray}
\left( \begin{array}{c}
\psi_{n+1\uparrow} \\
\psi_{n+1\downarrow} \\
\psi_{n\uparrow} \\
\psi_{n\downarrow} \end{array} \right) &=&
\left( \begin{array}{cccc}
\frac{E-\epsilon_n + h_n cos\theta_n}{t} & \frac{h_n sin\theta_n 
e^{-i\phi_n}} {t} & -1 & 0 \\
\frac{h_n sin\theta_n e^{i\phi_n}}{t} & \frac{E-\epsilon_n-h_n 
cos\theta_n}{t} & 0 & -1 \\
1 & 0 & 0 & 0 \\
0 & 1 & 0 & 0 \end{array} \right) \nonumber \\
 & & \times \left( \begin{array}{c}
\psi_{n\uparrow} \\
\psi_{n\downarrow} \\
\psi_{n-1\uparrow} \\
\psi_{n-1\downarrow} \end{array} \right)
\label{equation63}
\end{eqnarray}
So the total transfer matrix for the whole MQD device can be written as,
$P = \prod\limits_{l=N}^1 P_{l}$ \\
where, 
\begin{equation}
P_n=\left( \begin{array}{cccc}
\frac{E-\epsilon_n + h_n cos\theta_n}{t} & \frac{h_n sin\theta_n 
e^{-i\phi_n}} {t} & -1 & 0 \\
\frac{h_n sin\theta_n e^{i\phi_n}}{t} & \frac{E-\epsilon_n-h_n 
cos\theta_n}{t} & 0 & -1 \\
1 & 0 & 0 & 0 \\
0 & 1 & 0 & 0 \end{array} \right)
\label{equation64}
\end{equation}
For the whole lead-MQD-lead system, the transfer matrix $\mathcal{T}$ 
that relates the wave amplitude from the left lead ($-1$th and $0$th 
sites) to the right lead ($(N+1)$th and $(N+2)$th sites) is given by, 
\begin{equation}
\left( \begin{array}{c}
\psi_{N+2\uparrow} \\
\psi_{N+2\downarrow} \\
\psi_{N+1\uparrow} \\
\psi_{N+1\downarrow} \end{array} \right) = \mathcal{T}
\left( \begin{array}{c}
\psi_{0\uparrow} \\
\psi_{0\downarrow} \\
\psi_{-1\uparrow} \\
\psi_{-1\downarrow} \end{array} \right)
\label{equation65}
\end{equation}
where,
\begin{equation}
\mathcal{T} = M_R.P.M_L
\label{equation66}
\end{equation}
Here, $M_R$ is the transfer matrix for right boundary, i.e., $(N+1)$th site. 
It relates the wave amplitudes from $N$th site (in the MQD device) to 
$(N+1)$th site (in the right lead). Similarly, $M_L$ represents the transfer 
matrix for left boundary, i.e., $0$th site. It relates the wave amplitudes 
from $1$st site (in the MQD device) to $-1$th site (in the left lead).

\subsection{To calculate $M_L$ and $M_R$}

Let us calculate $M_L$ first. For the left (or right) lead, 
$\vec h.\vec \sigma = 0$. So the Hamiltonian for the lead (left or right) 
can be simplified as,
$H_{\mbox{lead}} = H_I + H_{II}$, where,
\begin{equation}
H_I = \sum_n \left(|n\uparrow \rangle \epsilon_{n\uparrow} 
\langle n\uparrow| + |n\downarrow \rangle \epsilon_{n\downarrow}
 \langle \downarrow|\right)
\label{equation67}
\end{equation}
and
\begin{eqnarray}
H_{II} & = & \sum_n \left(|n\uparrow \rangle t \langle n+1,\uparrow|
+|n\downarrow \rangle t \langle n+1,\downarrow| \right.\nonumber \\ 
& & \left. + |n+1,\uparrow \rangle t \langle n 
\uparrow| + |n+1,\downarrow \rangle t \langle n \downarrow|\right)
\label{equation68}
\end{eqnarray}
We set $\epsilon_n=\epsilon_0$ for all $n$ in the leads. Now, from the 
time independent Schr\"{o}dinger equation, operating $H$ on 
$|\phi \rangle$ we get two equations for $0$th site, relating the wave 
amplitudes of $1$st and $-1$th sites, as given below.
\begin{eqnarray}
\left(E-\epsilon_0 \right)\psi_{0\uparrow} & = & t_{LD}\psi_{1\uparrow} 
+ t_L\psi_{-1\uparrow} \nonumber \\
\left(E-\epsilon_0\right)\psi_{0\downarrow} & = & t_{LD}\psi_{1\downarrow} 
+ t_L\psi_{-1\downarrow} 
\label{equation69}
\end{eqnarray}
Now in the lead, according to the tight-binding model,
\begin{equation}
\psi_n = A e^{ikna}
\label{equation70}
\end{equation}
So $\psi_{-1}$ can be written in terms of $\psi_0$ as,
\begin{equation}
\psi_{-1} = \psi_0 e^{-i\beta_L}
\label{equation71}
\end{equation}
where, $\beta_L = ka$ and 
\begin{equation}
E = \epsilon_0 + 2t_{L(R)} cos\beta_{L(R)}
\label{equation72}
\end{equation}
From Eqs.~\ref{equation69}, \ref{equation70} and \ref{equation71}, 
$\psi_1$ can be expressed in terms of $\psi_0$ as,
\begin{equation}
\psi_1 = \left(\frac{t_L}{t_{LD}}e^{i\beta_L}\right)\psi_0
\label{equation73}
\end{equation}
Hence, the transfer matrix matrix equation for the $0$th site is of the
form,
\begin{equation}
\left( \begin{array}{c}
\psi_{1\uparrow} \\
\psi_{1\downarrow} \\
\psi_{0\uparrow} \\
\psi_{0\downarrow} \end{array} \right) = 
\left( \begin{array}{cccc}
\frac{t_L}{t_{LD}}e^{i\beta_L} & 0 & 0 & 0 \\
0 & \frac{t_L}{t_{LD}}e^{i\beta_L} & 0 & 0 \\
0 & 0 & e^{i\beta_L} & 0 \\
0 & 0 & 0 & e^{i\beta_L} \end{array} \right)
\left( \begin{array}{c}
\psi_{0\uparrow} \\
\psi_{0\downarrow} \\
\psi_{-1\uparrow} \\
\psi_{-1\downarrow} \end{array} \right)
\label{equation74}
\end{equation}
Similarly, from the time independent Schr\"{o}dinger equation, 
operating $H$ on $|\phi \rangle$ we get two equations for $(N+1)$th 
site, relating the wave amplitudes of $N$th and $(N+2)$th sites as 
given below.
\begin{eqnarray}
(E-\epsilon_0)\psi_{N+1\uparrow} & = & t_{R}\psi_{N+2\uparrow} 
+ t_{DR}\psi_{N-1\uparrow} \nonumber \\
(E-\epsilon_0)\psi_{N+1\downarrow} & = & t_R\psi_{N+2\downarrow} 
+ t_{DR}\psi_{N-1\downarrow} 
\label{equation75}
\end{eqnarray}
Exactly as before the transfer matrix matrix equation for the $(N+1)$th 
site can be written as, 
\begin{equation}
\left( \begin{array}{c}
\psi_{N+2\uparrow} \\
\psi_{N+2\downarrow} \\
\psi_{N+1\uparrow} \\
\psi_{N+1\downarrow} \end{array} \right) = 
\left( \begin{array}{cccc}
e^{i\beta_R} & 0 & 0 & 0 \\
0 & e^{i\beta_R} & 0 & 0 \\
0 & 0 & \frac{t_{DR}}{t_R}e^{i\beta_L} & 0 \\
0 & 0 & 0 & \frac{t_{DR}}{t_R}e^{i\beta_L} \end{array} \right)
\left( \begin{array}{c}
\psi_{N+1\uparrow} \\
\psi_{N+1\downarrow} \\
\psi_{N\uparrow} \\
\psi_{N\downarrow} \end{array} \right)
\label{equation76}
\end{equation}
Therefore,
\begin{eqnarray}
M_L & = & \mbox{the transfer matrix for the left boundary} \nonumber \\
& = & \left( \begin{array}{cccc}
\frac{t_L}{t_{LD}}e^{i\beta_L} & 0 & 0 & 0 \\
0 & \frac{t_L}{t_{LD}}e^{i\beta_L} & 0 & 0 \\
0 & 0 & e^{i\beta_L} & 0 \\
0 & 0 & 0 & e^{i\beta_L} \end{array} \right)\\ 
\label{equation77}
M_R & = & \mbox{the transfer matrix for the right boundary} \nonumber \\
& = & \left( \begin{array}{cccc}
e^{i\beta_R} & 0 & 0 & 0 \\
0 & e^{i\beta_R} & 0 & 0 \\
0 & 0 & \frac{t_{DR}}{t_R}e^{i\beta_L} & 0 \\
0 & 0 & 0 & \frac{t_{DR}}{t_R}e^{i\beta_L} \end{array} \right)
\label{equation78}
\end{eqnarray}

\subsection{To calculate the transmission probabilities of up and down 
spin electrons}

Thus the transfer matrix for the whole system (lead-MQD-lead) can be 
written as,
\begin{eqnarray}
\mathcal{T}= M_R.P.M_L
\label{equation79}
\end{eqnarray}
In order to calculate the transmission coefficient for the incident 
electrons with up or down spin using the transfer matrix method, the 
wave function amplitudes (Wannier amplitudes) have to be specified 
on proper atomic sites. 
\vskip 0.2in
\noindent
{\underline{Case 1: Up spin incidence from the left lead}} \\
\vskip 0.01cm
\noindent
The eigenvalue equation involving the transfer matrix relating the Wannier
amplitudes from sites $(N+2)$ and $(N+1)$ to sites $0$ and $-1$ is given by 
Eq.~\ref{equation65}.\\
\noindent
Let,\\
$\rho^{\uparrow \uparrow}$ = Reflection amplitude for up spin $(\uparrow)$ 
 reflected as  up spin $(\uparrow)$.\\
$\rho^{\uparrow \downarrow}$ = Reflection amplitude for up spin $(\uparrow)$
reflected as down spin $(\downarrow)$.\\
$\tau^{\uparrow \uparrow}$ = Transmission amplitude for up spin $(\uparrow)$
transmitted as up spin $(\uparrow)$. \\
$\tau^{\uparrow \downarrow}$ = Transmission amplitude for up spin $(\uparrow)$
transmitted as down spin $(\downarrow)$.\\
\noindent
Now, for the left lead, the wave amplitudes at the sites 0 and -1 can be 
written as,
\begin{eqnarray}
\psi_{-1\uparrow} & = & e^{-i\beta_L} + \rho^{\uparrow \uparrow}e^{i\beta_L} 
\nonumber\\
\psi_{-1\downarrow} & = & \rho^{\uparrow \downarrow}e^{i\beta_L} \nonumber \\
\psi_{0\uparrow} & = & 1+\rho^{\uparrow \uparrow} \nonumber \\
\psi_{0\downarrow} & = & \rho^{\uparrow \downarrow} 
\label{equation80}
\end{eqnarray}
Similarly, for the right lead, the wave amplitudes at the sites $(N+1)$ 
and $(N+2)$ become,
\begin{eqnarray}
\psi_{N+2\uparrow} = \tau^{\uparrow \uparrow}e^{i(N+2)\beta_R} \nonumber \\
\psi_{N+2\downarrow} = \tau^{\uparrow \downarrow}e^{i(N+2)\beta_R} \nonumber\\
\psi_{N+1\uparrow} = \tau^{\uparrow \uparrow}e^{i(N+1)\beta_R} \nonumber \\
\psi_{N+1\downarrow} = \tau^{\uparrow \downarrow}e^{i(N+1)\beta_R} 
\label{equation81}
\end{eqnarray}
Therefore, the transfer matrix eigenvalue equation can be rewritten 
in terms of these wave function amplitudes as, 
\begin{equation}
\left( \begin{array}{c}
\tau^{\uparrow \uparrow}e^{i(N+2)\beta_R} \\
\tau^{\uparrow \downarrow}e^{i(N+2)\beta_R} \\
\tau^{\uparrow \uparrow}e^{i(N+1)\beta_R} \\
\tau^{\uparrow \downarrow}e^{i(N+1)\beta_R} \end{array} \right) = \mathcal{T}.
\left( \begin{array}{c}
1+\rho^{\uparrow \uparrow} \\
\rho^{\uparrow \downarrow} \\
e^{-i\beta_L} + \rho^{\uparrow \uparrow}e^{i\beta_L}\\
\rho^{\uparrow \downarrow}e^{i\beta_L} \end{array} \right)
\label{equation82}
\end{equation}
Solving the above equation we can get the values of 
$\tau^{\uparrow \uparrow}$ and $\tau^{\uparrow \downarrow}$. The 
transmission coefficients $T^{\uparrow \uparrow}$ and 
$T^{\uparrow \downarrow}$ are defined as the ratio of the transmitted 
flux to the incident flux as,
\begin{eqnarray}
T_{\uparrow \uparrow} & = & \frac{t_R \sin\beta_R}{t_L \sin\beta_L} 
|\tau^{\uparrow \uparrow}|^2 \nonumber \\
T_{\uparrow \downarrow} & = & \frac{t_R \sin\beta_R}{t_L \sin\beta_L} 
|\tau^{\uparrow \downarrow}|^2 
\label{equation83}
\end{eqnarray}
Therefore, the total transmission probability for spin up is,
\begin{equation}
T_\uparrow = T_{\uparrow \uparrow} + T_{\uparrow \downarrow}
\label{equation84}
\end{equation}
\noindent
{\underline {Case 2: Down spin incidence from the left lead}} \\
\vskip 0.01 in
\noindent
Similarly, for a down spin incidence from the left lead, the 
amplitudes at sites $-1$, $0$, $(N+1)$ and $(N+2)$ are given by,
\begin{eqnarray}
\psi_{0\uparrow} & = & \rho^{\downarrow \uparrow} \nonumber \\
\psi_{0\downarrow} & = & 1+\rho^{\downarrow \downarrow} \nonumber \\
\psi_{-1\uparrow} & = & \rho^{\downarrow \uparrow}e^{i\beta_L} \nonumber \\
\psi_{-1\downarrow} & = & e^{-i\beta_L} + \rho^{\downarrow \downarrow}
e^{i\beta_L}
\label{equation85}
\end{eqnarray} 
and
\begin{eqnarray}
\psi_{N+2\uparrow} & = & \tau^{\downarrow \uparrow}e^{i(N+2)\beta_R} 
\nonumber \\
\psi_{N+2\downarrow} & = & \tau^{\downarrow \downarrow}e^{i(N+2)\beta_R} 
\nonumber \\
\psi_{N+1\uparrow} & = & \tau^{\downarrow \uparrow}e^{i(N+1)\beta_R} 
\nonumber \\
\psi_{N+1\downarrow} & = & \tau^{\downarrow \downarrow}e^{i(N+1)beta_R} 
\label{equation86}
\end{eqnarray}
So, as before, the transfer matrix equation for down spin incidence can be
rewritten in terms of these wave amplitudes as, \\
\begin{equation}
\left( \begin{array}{c}
\tau^{\downarrow \uparrow}e^{i(N+2)\beta_R} \\
\tau^{\downarrow \downarrow}e^{i(N+2)\beta_R} \\
\tau^{\downarrow \uparrow}e^{i(N+1)\beta_R} \\
\tau^{\downarrow \downarrow}e^{i(N+1)\beta_R} \end{array} \right)=\mathcal{T}.
\left( \begin{array}{c}
\rho^{\downarrow \uparrow} \\
1+\rho^{\downarrow \downarrow} \\
\rho^{\downarrow \uparrow}e^{i\beta_L} \\
e^{-i\beta_L} + \rho^{\downarrow \downarrow}e^{i\beta_L} \end{array} \right)
\label{equation87}
\end{equation}
This equation can be solved to compute $\tau^{\downarrow \downarrow}$ 
and $\tau^{\downarrow \uparrow}$. The transmission coefficients 
$T^{\downarrow \downarrow}$ and
$T^{\downarrow \uparrow}$ are defined by the ratio of the transmitted 
flux to the incident flux as, \\
\begin{eqnarray}
T_{\downarrow \downarrow} & = & \frac{t_R \sin\beta_R}{t_L \sin\beta_L} 
|\tau^{\downarrow \downarrow}|^2 \nonumber \\
T_{\downarrow \uparrow} & = & \frac{t_R \sin\beta_R}{t_L \sin\beta_L} 
|\tau^{\downarrow \uparrow}|^2 
\label{equation88}
\end{eqnarray}
Therefore, the total transmission probability for spin down is, \\
\begin{equation}
T_\downarrow = T_{\downarrow \downarrow} + T_{\downarrow \uparrow}
\label{equation89}
\end{equation}
At low temperatures, the transport of such mesoscopic systems is 
ballistic and coherent. In this regime, transmitted spin dependent 
current flowing through the interacting region with arbitrary magnetic 
configurations, symmetrically attached to two NM leads is given 
by the relation~\cite{r29,r30,r35},
\begin{equation}
I_{\sigma \sigma^{\prime}} (V)= \frac{e}{h} \int 
\left(f_S-f_D\right) T_{\sigma \sigma^{\prime}}(E)~dE
\label{equation90}
\end{equation}
where, $f_{S(D)}=f\left(E-\mu_{S(D)}\right)$ gives the Fermi distribution
function with the electrochemical potential $\mu_{S(D)}=E_F\pm eV/2$,
$E_F$ being the equilibrium Fermi energy.

\section{Numerical results and discussion}

Based on the above theoretical framework now we present our numerical results
for NM/MQD/NM heterostructure. We assume that the MQD device is constructed
by repeating a specific unit cell which is made up of some magnetic (A) and 
non-magnetic (B) atoms having different magnetic moment orientations. By 
applying an external magnetic field, the magnetization direction of the 
NM/MQD/NM system can be switched. Four different unit cell configurations
are taken into account to reveal different aspects of spin dependent 
transport through a quantum heterostructure, and these results are 
critically analyzed below.

Before presenting the numerical results let us first mention the values of
different parameters those are considered throughout our numerical 
calculations. For the sake of simplicity, the on-site energies both in the
two leads and in MQD device are taken to be zero. The nearest-neighbor 
hopping integral $t_{n,n+1}$ ($=t$) in the MQD device is set to $1$, and, 
in the two side-attached leads the hopping strength $t_{L(R)}$ is also 
fixed at $1$. The spin flip parameter is chosen as $h=1$. To narrate the 
coupling effect~\cite{r31,r32,r34,r344} we focus our results for the two 
limiting cases
depending on the coupling strength between the MQD device to the source and
drain leads. In one case we call it weak-coupling limit which is described
by the condition $t_{LD(RD)} << t$. For this regime we choose
$t_{LD}=t_{RD}=0.2$. While the other case, called as strong-coupling limit,
is defined by the condition $t_{LD(RD)} \sim t$, and here we set
$t_{LD}=t_{RD}=0.8$. Throughout the analysis we choose the units $c=e=h=1$, 
for simplification, and set the electronic temperature of the system to 
zero.
 
\begin{figure}[ht]
{\centering \resizebox*{8cm}{7cm}{\includegraphics{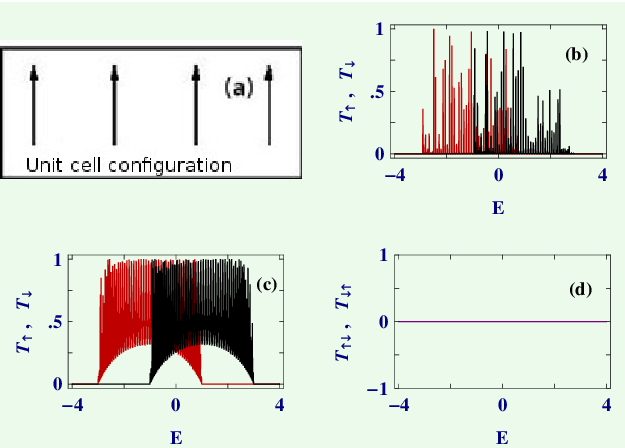}}\par}
\caption{(Color online). Spin dependent transmission probabilities as a 
function of injecting electron energy $E$ for a $64$-site MQD device
considering the unit cell configuration as given in (a). The unit cell is
made up of four magnetic atoms $A$, each having parallel magnetic moment.
The net up ($T_{\uparrow}$, red color) and down ($T_{\downarrow}$, black 
color) spin transmission probabilities for the weak-coupling limit are 
shown in (b), while these probabilities for the strong-coupling limit 
are given in (c). Spin flip transmission probabilities 
($T_{\uparrow\downarrow}$ and $T_{\downarrow\uparrow}$) for both the
weak- and strong-coupling limits are presented in (d).}
\label{config1}
\end{figure}

Figure~\ref{config1} presents the spin dependent transmission probabilities
as a function of injecting electron energy $E$ for a specific unit cell
configuration shown in (a), where the unit cell is made up of four $A$-type
(magnetic) atoms and the moment of each atom is aligned along positive 
$Z$-direction. The net up and down spin probabilities ($T_{\uparrow}$ and
$T_{\downarrow}$) for both the weak- and strong-coupling limits are 
computed and they are presented in (b) and (c), respectively, while in 
(d) we present spin flip transmission probabilities ($T_{\uparrow\downarrow}$ 
and $T_{\downarrow\uparrow}$). From the spectra it is observed that the
transmission probability exhibits sharp resonant peaks in the limit of 
weak-coupling, while these peaks get broadened and achieve much higher 
value in the limit of strong-coupling. The contribution for the 
broadening of the resonant peaks appears from the broadening of energy
levels of the MQD device in this strong-coupling limit. All these resonant
peaks are associated with the energy eigenvalues of the bridging system, 
and thus, we can say that the transmission-energy spectrum manifests itself 
the electronic structure of the device sandwiched between two electrodes. 
Most interestingly we see from the spectra given in Figs.~\ref{config1}(b) 
and (c) that the MQD device can exhibit spin polarization for a wide range 
of energy. There is sharp energy shift between the up and down spin 
transmission spectra. For one energy region ($-3\leq E \leq -1$) only up
spin electrons, while for the other region ($1\leq E \leq 3$) 
only down spin electrons are allowed to move from source to drain through 
the MQD device. Within the range $-1\leq E \leq 1$ the overlap between two 
transmission probabilities takes place. Thus setting the Fermi energy 
to a suitable energy zone ($-3\leq E \leq -1$ or $1\leq E \leq 3$) we 
can get either up or down spin transmission across the junction which 
yields a selective spin transmission. This is essentially what we expect 
from our model. For this model where all magnetic moments are aligned 
along $Z$-direction, the spin-flip transmission
\begin{figure}[ht]
{\centering \resizebox*{8cm}{7cm}{\includegraphics{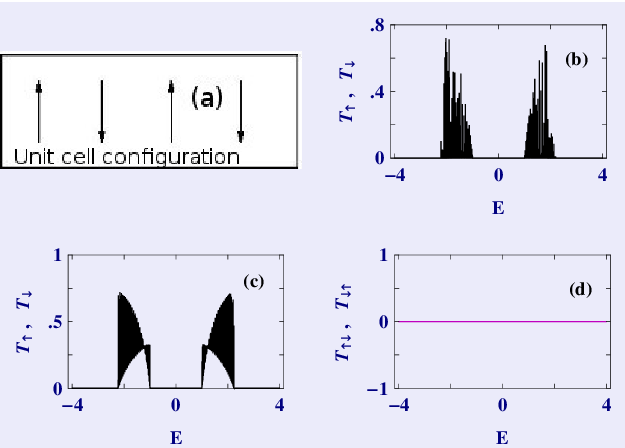}}\par}
\caption{(Color online). Spin dependent transmission probabilities as a 
function of energy for a $64$-site MQD device where the unit cell, 
shown in (a), is made up of four magnetic atoms $A$, having anti-parallel 
magnetic moment. The spectra (b), (c) and (d) correspond to the identical 
meanings as described in Fig.~\ref{config1}.}
\label{config2}
\end{figure}
\begin{figure}[ht]
{\centering \resizebox*{8cm}{7cm}{\includegraphics{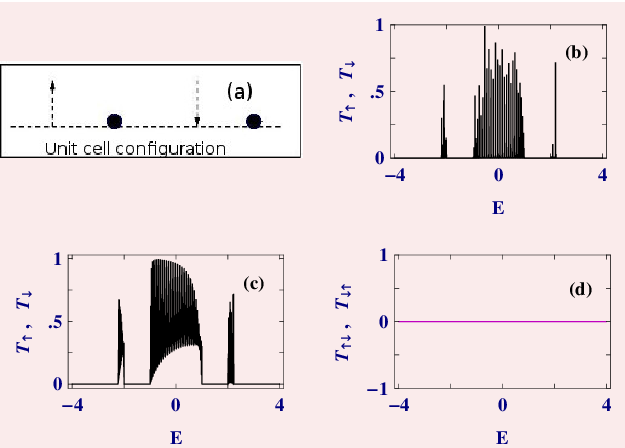}}\par}
\caption{(Color online). Spin dependent transmission probabilities as a 
function of energy for a $64$-site MQD device where the unit cell, 
shown in (a), is made up of two magnetic atoms $A$, oriented in 
anti-parallel configuration, and two non-magnetic atoms $B$. The spectra 
(b), (c) and (d) correspond to the identical meanings as described in 
Fig.~\ref{config1}.}
\label{config3}
\end{figure}
probability becomes zero for the entire energy region (see 
Fig.~\ref{config1}(d)). This can be explained through a simple mathematical
argument. The term $\vec{\bf{h}}_n.\vec{\bf{\sigma}}$ in the Hamiltonian 
$H_D$ is responsible for spin flipping. Since all the moments are aligned
along $Z$-direction, this terms simplifies to $h_n\sigma_z$, which is
\begin{figure}[ht]
{\centering \resizebox*{5cm}{2cm}{\includegraphics{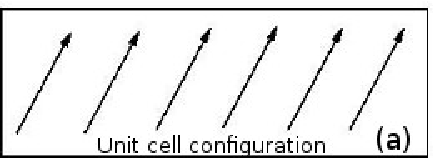}}
\vskip 0.2cm
\centering \resizebox*{8cm}{7cm}{\includegraphics{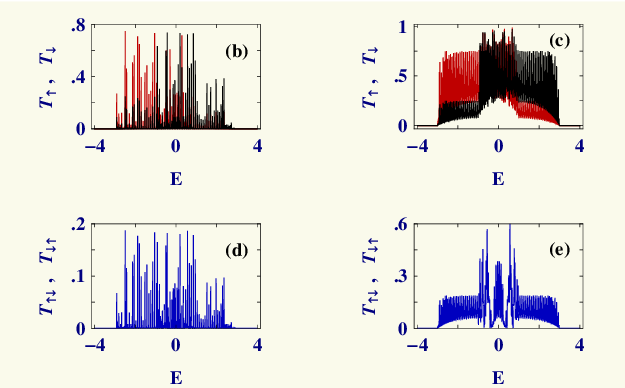}}\par}
\caption{(Color online). Spin dependent transmission probabilities as a 
function of energy for a $64$-site MQD device where the unit cell, 
shown in (a), is made up of magnetic atoms $A$ those are oriented at an 
angle $\theta$ with respect to $Z$-axis. The net up ($T_{\uparrow}$, red 
color) and down ($T_{\downarrow}$, black color) spin transmission 
probabilities for the weak- and strong-coupling limits are shown in (b) and
(c), respectively, while the spin-flip transmission probabilities
($T_{\uparrow\downarrow}$ and $T_{\downarrow\uparrow}$, they exactly 
overlap with each other) for these coupling limits are shown in (d) and 
(e), respectively. Here $\theta$ is fixed at $\pi/3$.}
\label{config4}
\end{figure}
\begin{figure}[ht]
{\centering \resizebox*{7cm}{3.5cm}{\includegraphics{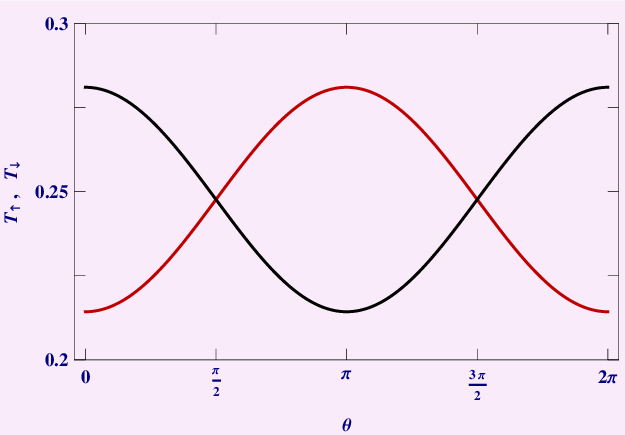}}\par}
\caption{(Color online). $T_{\uparrow}$ (red curve) and 
$T_{\downarrow}$ (black curve) as a function of $\theta$ for a $64$-site 
MQD (all sites are magnetic) device in the strong-coupling limit for the 
typical energy $E=0.25$.}
\label{tht}
\end{figure}
diagonal and does not contain spin-flip factors $\sigma_+$ and $\sigma_-$, 
associated with $\sigma_x$ and $\sigma_y$. Therefore, no spin-flip 
transmission is obtained for this configuration.

Now move to the device where all sites are magnetic ($A$-type), but the
moments are aligned in a perfect anti-parallel configuration as shown in
Fig.~\ref{config2}(a). The results for such a quantum device with $64$
sites are shown in Figs.~\ref{config2}(b)-(d), where different
spectra correspond to the identical meanings as described in 
Fig.~\ref{config1}. Quite interestingly we notice that for this 
configuration both the up and down spin bands overlap with each other,
and accordingly, spin polarization cannot be achieved. This band overlap
can be easily understood from the Hamiltonian $H_D$ since up and down 
spin electrons experience exactly identical potential while traversing 
through the MQD device, and thus, the energy spectra for these two spin
\begin{figure}[ht]
{\centering \resizebox*{5cm}{2cm}{\includegraphics{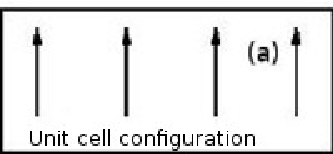}}
\vskip 0.2cm
\centering \resizebox*{7.5cm}{7.5cm}{\includegraphics{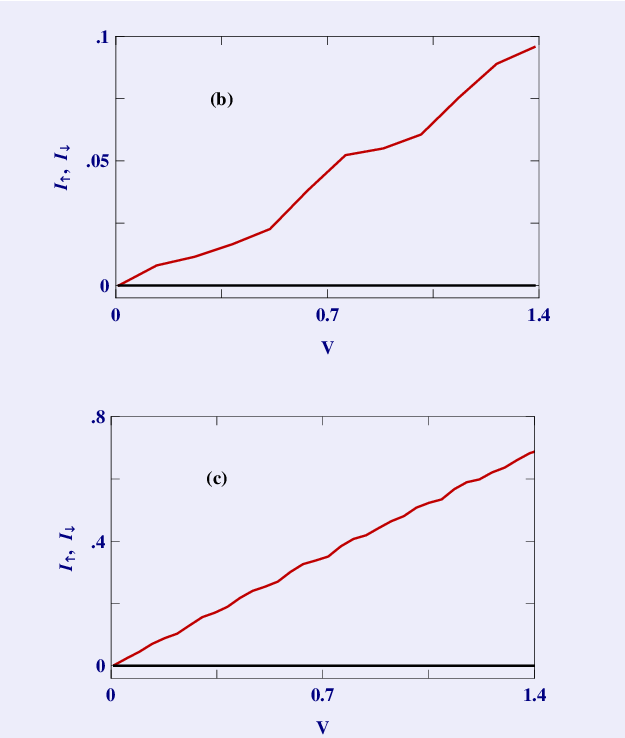}}\par}
\caption{(Color online). Spin dependent currents ($I_{\uparrow}$ and 
$I_{\downarrow}$) as 
a function of bias voltage $V$ for a $64$-site MQD device considering the
unit cell where all moments are aligned in a perfect parallel configuration
(shown in (a)), where the red and black lines correspond to $I_{\uparrow}$ 
and $I_{\downarrow}$, respectively. (b) Weak-coupling limit and 
(c) strong-coupling limit. Here we set the Fermi energy $E_F=-2.0$.}
\label{curr1}
\end{figure}
cases are identical which yield identical transmission-energy spectra.
Along with this we also find a gap in the spectrum across $E=0$. This is
exactly what we get in a binary lattice model since for this configuration
successive moments are aligned in anti-parallel configuration providing an
equivalent binary lattice geometry. Like Fig.~\ref{config1}, here also we
get vanishing spin flip transmission (Fig.~\ref{config2}(d)) as $\theta$ 
becomes $0$ or $\pi$ for the moments.

In Fig.~\ref{config3} the results are given for another MQD device where 
two magnetic moments are separated by a non-magnetic one and the moments
are oriented in perfect anti-parallel configuration. Similar to the previous
case (i.e., Fig.~\ref{config2}) here also we do not get any separated
spin channel and hence no spin polarization will be available. For this
device the spectrum exhibits two finite gaps at the energy band edges
associated with unit cell configuration, and like previous systems, spin 
flipping does not take place.

Finally, focus on the MQD device where all the sites are magnetic and they
are oriented at a particular angle $\theta$ with respect to $Z$-axis. The
unit cell configuration for such a system is schematically shown in 
Fig.~\ref{config4}(a) and the corresponding results are placed in 
Fig.~\ref{config4}(b)-(e). Here we set $\theta=\pi/3$. For this set-up
spin polarization is achieved for a wide range of energy $E$, like 
the first configuration (Fig.~\ref{config1}), which is more clearly seen
from the result of the strong-coupling case (Fig.~\ref{config4}(c)).
Though the present configuration exhibits spin polarization for a large
energy window, the degree of spin polarization is less compared to the
\begin{figure}[ht]
{\centering \resizebox*{5cm}{2cm}{\includegraphics{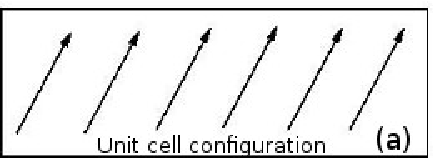}}
\vskip 0.2cm
\centering \resizebox*{7.5cm}{7.5cm}{\includegraphics{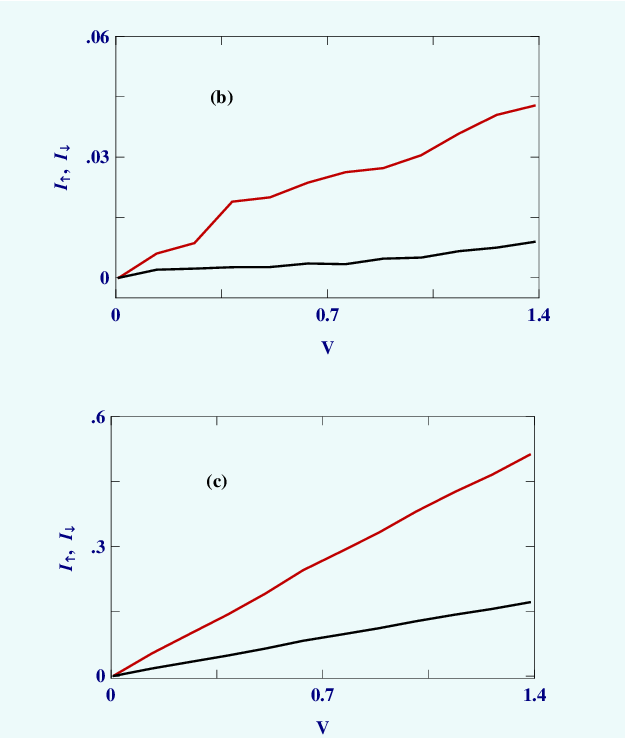}}\par}
\caption{(Color online). Spin dependent currents ($I_{\uparrow}$ and 
$I_{\downarrow}$)
as a function of voltage $V$ for a $64$-site MQD device where all the 
magnetic moments are aligned along a specific direction (as shown in (a)) 
making an angle $\theta$ with respect to $Z$-axis, where the red and 
black lines correspond to $I_{\uparrow}$ and $I_{\downarrow}$, 
respectively. (a) Weak-coupling limit and (b) strong-coupling limit.
Here we choose $E_F=-2.0$ and $\theta=\pi/3$.}
\label{curr4}
\end{figure}
device where all the moments are aligned in a perfect parallel 
configuration (Fig.~\ref{config1}(a)). In the later case a perfect separation
takes place between up and down spin channels yielding $100\%$ spin 
polarization, while for the device where moments are aligned in a particular 
direction making an angle $\theta$, an overlap between two channels takes 
place though the transmission amplitudes are different providing less
degree of polarization. In addition we also find that for this MQD device
spin-flip transmission is obtained (see Figs.~\ref{config4}(d) and (e))
and the appearance of spin-flip transmission probabilities can be easily
understood from the spin-flip term $\vec{\bf h}_n.\vec{\bf\sigma}$ as
in contains $\sigma_+$ and $\sigma_-$ through the components $\sigma_x$ 
and $\sigma_y$ which are responsible for spin flipping. Obviously we 
should expect more flipping with increasing the orientation angle $\theta$.

So now the question naturally comes how the net up and down 
spin probabilities get changed with $\theta$. To answer it in Fig.~\ref{tht}
we present the results of $T_{\uparrow}$ (red curve) and $T_{\downarrow}$
(black curve) as a function of $\theta$ for a MQD device, considering
all sites are magnetic, at a typical energy $E=0.25$. It is interesting
to see that for $\theta=0$ the difference between $T_{\uparrow}$ and
$T_{\downarrow}$ becomes maximum (viz, maximum polarization) and it 
(the difference) gradually decreases to zero at $\theta=\pi/2$ (viz, no 
spin polarization). This difference
starts increasing with the further increment of $\theta$ and eventually 
reaches to a maximum for $\theta=\pi$, reversing the sign of spin 
polarization compared to $\theta=0$. Thus a periodic variation is naturally
expected as function of $\theta$ and it is also clearly reflected from 
the spectra.

All the basic features of spin dependent transmission probabilities together 
with spin polarization studied above become much more clearly visible from
\begin{figure}[ht]
{\centering \resizebox*{5cm}{2cm}{\includegraphics{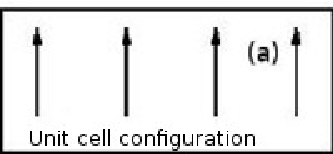}}
\vskip 0.2cm
\centering \resizebox*{7.5cm}{7.5cm}{\includegraphics{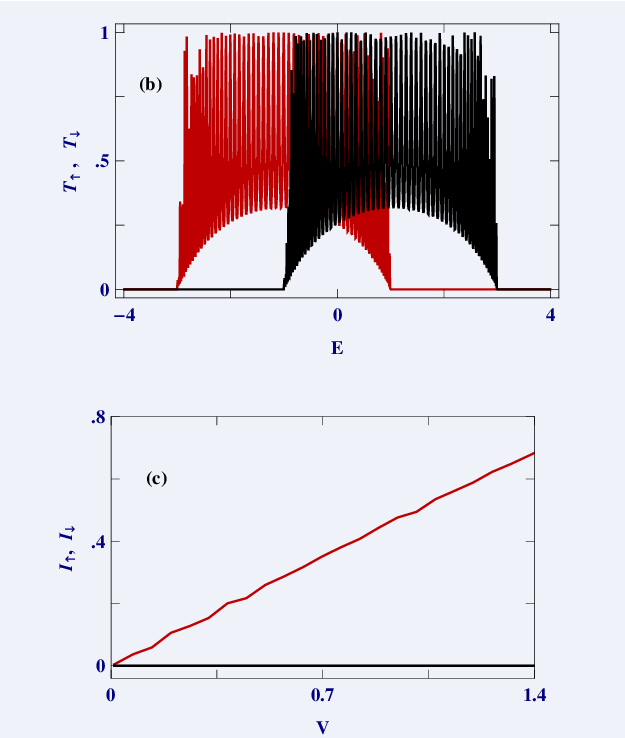}}\par}
\caption{(Color online). Spin dependent transmission probabilities and 
corresponding currents for a MQD device considering odd number of magnetic 
sites ($N=65$) where all the magnetic moments are aligned in a perfect 
parallel configuration. All the results are computed for the strong-coupling 
limit where the curves with red and black colors represent the up and down 
spin cases, respectively. We set $E_F=-2.0$.}
\label{trancurr}
\end{figure}
current-voltage characteristics. The current through the bridge system can 
be determined by integrating spin dependent transmission function over a
suitable energy window, as prescribed in Eq.~\ref{equation90}. Let us 
start with Fig.~\ref{curr1} where net up (red line) and down (black line) 
spin currents are shown as a function of bias voltage $V$ for a $64$-site
MQD device in which all the magnetic moments are aligned in a perfect
parallel configuration (viz, $\theta=0$). The results for the two different 
coupling cases are shown in (b) and (c), where all these currents are 
computed setting the Fermi energy $E_F=-2.0$ (the zone where only up spin 
electron can transmit, see Figs.~\ref{config1}(b) and (c)). In the limit of 
weak-coupling current provides step-like behavior, while it becomes quite
\begin{figure}[ht]
{\centering \resizebox*{5cm}{2cm}{\includegraphics{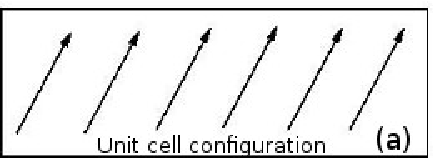}}
\vskip 0.2cm
\centering \resizebox*{7.5cm}{7.5cm}
{\includegraphics{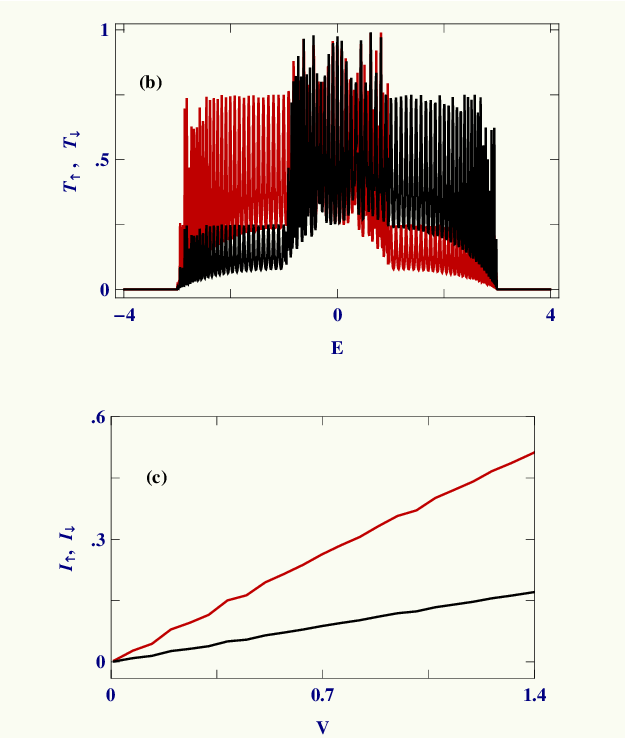}}\par}
\caption{(Color online). Spin dependent transmission probabilities and 
associated currents for a MQD device considering odd number of magnetic 
sites ($N=65$) where all the magnetic moments are aligned along a specific 
direction (as shown in (a)) making an angle $\theta=\pi/3$ with respect 
to $Z$-axis. The spectra shown in (b) and (c) correspond to the identical 
meanings as described in Fig.~\ref{trancurr} and they are computed for the 
same parameter values as taken in Fig.~\ref{trancurr}.}
\label{trancurrconfig4}
\end{figure}
continuous in the limit of strong-coupling. It is solely associated with
the transmission function as the current is computed by integrating this
function. From these spectra the behavior of spin polarization can be 
clearly examined. It is observed that for the entire voltage window only
up spin current (red line) is passing through the junction since the other
current (black line) drops exactly to zero. Certainly, for this wide
bias region the device exhibits $100\%$ spin polarization. So, the 
spin polarization efficiency defined by the condition
$\eta=(|I_{\uparrow}-I_{\downarrow}|)/(I_{\uparrow}+I_{\downarrow})$ 
becomes unity. Exactly opposite scenario i.e., finite down spin
current and vanishing up spin current across the junction, is obtained 
when the Fermi energy is fixed at $2.0$. Thus setting the Fermi energy
to a suitable energy zone we can get selective polarized spin current
with maximum efficiency $\eta=1$.

Since for the other two MQD devices, those are formed by repeating the unit 
cell configurations given in Figs.~\ref{config2}(a) and \ref{config3}(a), 
up and down spin channels are not separated in energy scale we cannot expect 
spin polarization,
\begin{figure}[ht]
{\centering \resizebox*{5cm}{2cm}{\includegraphics{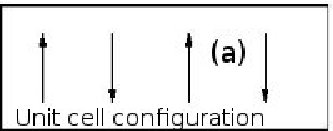}}
\vskip 0.2cm
\centering \resizebox*{7.5cm}{7.5cm}
{\includegraphics{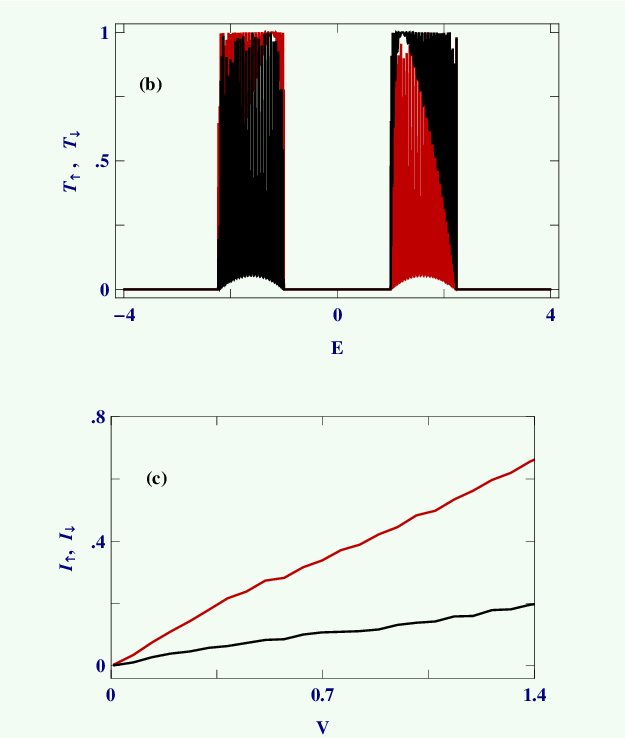}}\par}
\caption{(Color online). Spin dependent transmission probabilities and 
corresponding currents for a MQD device considering odd number of magnetic 
sites ($N=65$) where the magnetic moments are arranged in perfect 
anti-parallel configuration (as shown in (a)), where the spectra (b) and (c) 
represent the similar meanings as described in Fig.~\ref{trancurr} and 
they are computed for the identical parameter values as taken in 
Fig.~\ref{trancurr}. Here we assume that the magnetic moment at the end
site of the chain (viz, $n=65$) is aligned along positive $Z$-axis.}
\label{trancurrconfig2}
\end{figure}
and accordingly, we do not discuss the $I$-$V$ characteristics for these
systems to save space. So finally move to the system where all the moments
are aligned in a specific direction making an angle $\theta$ with respect 
to $Z$-direction. The results are shown in Fig.~\ref{curr4}. Like 
Fig.~\ref{curr1} here also current exhibits step-like 
and continuous-like nature of current in the limit of weak and strong 
coupling cases, respectively. But as $\theta$ is finite ($\ne 0$ or $\pi$),
a non-vanishing down spin current where the contribution comes from the spin
flipping is obtained together with up spin current though their amplitudes 
differ significantly yielding a finite polarization, which is obviously 
less then $100\%$ (viz, $\eta<1$). The reverse scenario will also be 
obtained when the Fermi level is placed at $2.0$, which is not shown here 
to save space.

The results studied so far are worked out for a typical system size
considering $N=64$ (even $N$). So one can think whether these features 
\begin{figure}[ht]
{\centering \resizebox*{5cm}{2cm}{\includegraphics{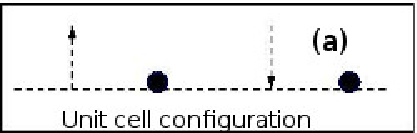}}
\vskip 0.2cm
\centering \resizebox*{7.5cm}{7.5cm}
{\includegraphics{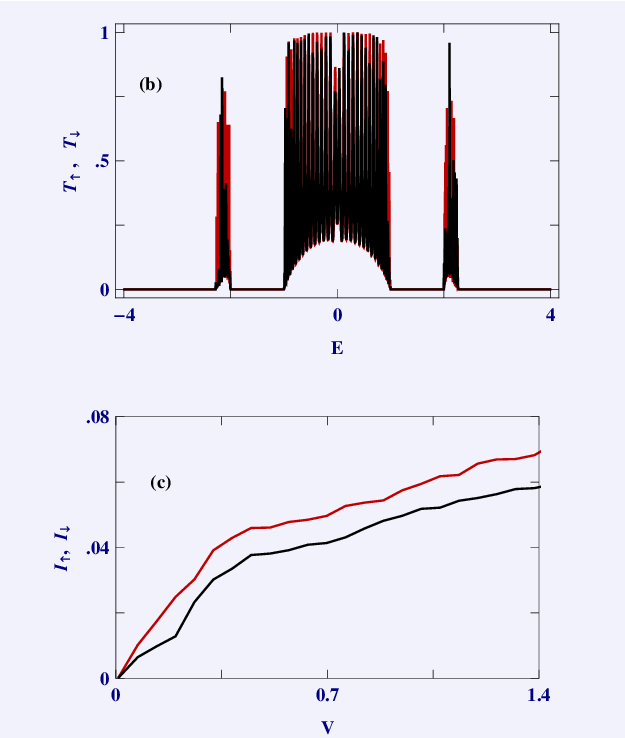}}\par}
\caption{(Color online). Spin dependent transmission probabilities and 
associated currents for a MQD device considering odd number of magnetic 
sites ($N=65$) where the magnetic moments are arranged as shown in (a),
and the other two spectra ((b) and (c)) represent the identical meanings 
as described in Fig.~\ref{trancurr}. The values of all physical parameters
remain same as considered in Fig.~\ref{trancurr}, except $E_F$ which is
set at $-2.15$. Here we also assume that the magnetic moment at the end
site of the chain (viz, $n=65$) is aligned along positive $Z$-axis, like
Fig.~\ref{trancurrconfig2}.}
\label{trancurrconfig3}
\end{figure}
get changed for such MQD devices when the system size ($N$) becomes odd or
remain same like what we get for the case of even $N$. To reveal this fact 
start with Fig.~\ref{trancurr} where we present the results for a $65$-site 
($N=65$) MQD device considering the perfect parallel alignment of all 
magnetic moments. In Fig.~\ref{trancurr}(b) we show the dependence of net 
up (red color) and down (black color) spin transmission probabilities, and 
the corresponding currents are placed in Fig.~\ref{trancurr}(c). Like 
Fig.~\ref{config1}(c) (where we set $N=64$) here also we find a clear 
separation between up and down spin transmission spectra 
(Fig.~\ref{trancurr}(b)), and the width of this energy separation is exactly 
identical to the previous case. This behavior is nicely reflected in the 
current-voltage characteristics (see Fig.~\ref{trancurr}(c)), similar
to Fig.~\ref{curr1}(c). Thus for this configuration, where all moments 
are aligned in a perfect parallel configuration, we expect identical 
behavior both for odd and even $N$.

In the same footing we can anticipate identical behavior for the MQD 
device where all the moments are arranged along a specific direction
(Fig.~\ref{trancurrconfig4}(a)) making an angle $\theta$ with respect to 
$Z$-axis. The results are shown in Fig.~\ref{trancurrconfig4} for a 
$65$-site magnetic system setting $\theta=\pi/3$, where all the other 
physical parameters kept unchanged as taken in Fig.~\ref{trancurr}. The
transmission probabilities for both up and down spin electrons 
(Fig.~\ref{trancurrconfig4}(b)) look exactly identical what we get in
Fig.~\ref{config4}(c) for a $64$-site system, and thus, spin dependent
currents (Fig.~\ref{trancurrconfig4}(c)) exhibit identical variation like
Fig.~\ref{curr4}(c). Therefore, for this configuration we also get identical
behavior irrespective of $N$.

But, some different features can be expected depending on even and odd $N$ 
for the other two configurations (Figs.~\ref{config2}(a) 
and \ref{config3}(a)). The reason is that the magnetic moment of the last
site breaks the symmetry between the potentials experienced by up and down
spin electrons. And accordingly a finite spin polarization is obtained,
though it becomes too small. The results are shown in 
Figs.~\ref{trancurrconfig2} and \ref{trancurrconfig3} considering $N=65$,
where we assume that the magnetic moment at the end site is aligned along
the positive $Z$-direction which essentially makes the difference between
up and down spin probabilities as well as currents. In both these two 
cases we get finite spin polarization as up spin current is higher than 
the other one (Figs.~\ref{trancurrconfig2}(c) and \ref{trancurrconfig3}(c)). 
Exactly opposite scenario (viz, higher down spin current compared to the 
up spin current) is obtained when the orientation of the magnetic moment 
placed at the end site gets flipped from up to down viz, oriented along 
negative $Z$-direction.

\section{Concluding remarks}

To conclude, in the present work we explore spin dependent 
transport phenomena through a one-dimensional quantum heterostructure 
composed of magnetic and non-magnetic quantum dots. A simple tight-binding
framework is given to describe the model where the heterostructure
is coupled to two one-dimensional non-magnetic electrodes, and all the
calculations are done based on transfer matrix method. Several cases
are analyzed depending on the configuration of the bridging system.
From our numerical results which describe two-terminal spin dependent
transmission probabilities along with transport currents, we show that
under certain condition {\em $100\%$ spin polarization can be achieved
for a wide range of bias voltage.} Certainly, this is an important 
observation and can be utilized to design an efficient spin filter 
device in nano-scale level. Apart from this our detailed theoretical 
analysis based on transfer matrix method can be implemented quite easily 
to any spin based quantum structure to describe transport phenomena.

Throughout the presentation, we have made several important assumptions. 
Below we discuss briefly some of these approximations.

$\bullet$ All the calculations have been worked out at absolute zero
temperature, but the physical picture studied above will not change at 
non-zero finite temperature as long as the thermal energy $(k_BT)$ is 
less than the average spacing of energy levels of the bridging material.

$\bullet$ The effect of electron-electron (e-e) correlation has not been 
taken into account. The inclusion of e-e correlation is a major challenge 
to us, since over the past few years much efforts have been made to 
incorporate this effect, but no proper theory has been well developed.

$\bullet$ Electron-phonon (e-ph) interaction has also been neglected in our
theoretical formulation, as we are dealing with zero temperature. But even 
at finite temperature one can neglect this effect as the broadening of 
energy levels due to e-ph interaction is much smaller than the broadening
caused by coupling of the MQD device with electrodes.

$\bullet$ All the results have been presented for ordered systems, but in 
presence of impurity transport properties can change which demands further
study and we hope it will be available in our next work.

\end{document}